\documentclass[12pt,a4paper]{article}

\RequirePackage[l2tabu, orthodox]{nag}

\usepackage{mjheppub}
\usepackage{lipsum}
\usepackage{amsthm,amssymb,amsmath,epic,eepic,float}
\usepackage{rotating,epsfig,indentfirst,array,varioref}
\usepackage{appendix,marginnote,bbm,tikz,pgf,mathtools}
\usepackage{setspace}

\usepackage{tikz}
\usetikzlibrary{decorations.pathreplacing}
\usetikzlibrary{calc}

\usepackage{color}

\let\emptyset\varnothing

\def\ll{\left\lgroup}
\def\rr{\right\rgroup}

\def\leq{\leqslant}
\def\geq{\geqslant}

\newcommand{\sll}{\mathrm{sl}}

\newcommand{\slN}{\mathfrak{sl}_N}
\newcommand{\slthree}{\mathfrak{sl}_3}
\newcommand\mydef{\stackrel{\mathclap{\normalfont\mbox{def}}}{=}}

\newdimen\tableauside\tableauside=1.0ex
\newdimen\tableaurule\tableaurule=0.4pt
\newdimen\tableaustep
\def\phantomhrule#1{\hbox{\vbox to0pt{\hrule height\tableaurule
width#1\vss}}}
\def\phantomvrule#1{\vbox{\hbox to0pt{\vrule width\tableaurule
height#1\hss}}}
\def\sqr{\vbox{%
  \phantomhrule\tableaustep

\hbox{\phantomvrule\tableaustep\kern\tableaustep\phantomvrule\tableaustep}%
  \hbox{\vbox{\phantomhrule\tableauside}\kern-\tableaurule}}}
\def\squares#1{\hbox{\count0=#1\noindent\loop\sqr
  \advance\count0 by-1 \ifnum\count0>0\repeat}}
\def\tableau#1{\vcenter{\offinterlineskip
  \tableaustep=\tableauside\advance\tableaustep by-\tableaurule
  \kern\normallineskip\hbox
    {\kern\normallineskip\vbox
      {\gettableau#1 0 }%
     \kern\normallineskip\kern\tableaurule}%
  \kern\normallineskip\kern\tableaurule}}
\def\gettableau#1 {\ifnum#1=0\let\next=\null\else
  \squares{#1}\let\next=\gettableau\fi\next}

\def\be{\begin{equation}}
\def\ee{\end{equation}}
\def\ba{\begin{array}}
\def\ea{\end{array}}

\restylefloat{figure}

\newcommand{\cB}{\mathcal{B}}

\newcommand{\cL}{\mathcal{L}}

\newcommand{\cT}{\mathcal{T}}

\newcommand{\cW}{\mathcal{W}}

\def\ll{ \left\lgroup}
\def\rr{\right\rgroup}

\begin{document}

\title{\boldmath 
Second level semi-degenerate fields in $\cW_3$ Toda theory: matrix element and differential equation
} 

\author{Vladimir Belavin  $^1$,}
\author{Xiangyu Cao $^2$,}
\author{Benoit Estienne   $^3$,}
\author{Raoul Santachiara $^2$ }
\affiliation{\vspace{2mm} $^1$ 
I E Tamm Department of Theoretical Physics, P N Lebedev Physical Institute,
Leninsky Avenue 53, 119991 Moscow, Russia, 
\newline
Department of Quantum Physics, Institute for Information Transmission Problems,
Bolshoy Karetny per. 19, 127994 Moscow, Russia
\newline
Moscow Institute of Physics and Technology, 
Dolgoprudnyi, 141700 Moscow region, Russia}
\affiliation{$^2$ LPTMS, CNRS (UMR 8626), Universit\'e Paris-Saclay, 15 rue Georges Cl\'emenceau, 91405 Orsay, 
France}
\affiliation{$^3$ LPTHE, CNRS, Sorbonne 
Universit\'es and Universit\'{e} Pierre et Marie Curie, 4 Place Jussieu, 75252 Paris Cedex 05, France}

\emailAdd{
belavin@lpi.ru, xiangyu.cao08@gmail.com, raoul.santachiara@u-psud.fr
}

\abstract{
In a recent study we considered $\cW_3$ Toda 4-point functions 
that involve matrix elements of a primary field with the highest-weight in the \textit{adjoint} 
representation of $\slthree$.  
We generalize this result by considering a semi-degenerate primary field, which has one null vector at level two. We obtain a sixth-order Fuchsian differential 
equation for the conformal blocks. We discuss the presence of multiplicities, the matrix elements and the fusion rules.
}

\keywords{
$\cW_N$ algebra, 2-dimensional conformal field theory, Fuchsian differential 
equations.
}

\maketitle
\flushbottom

%SECTION.01
\section{Introduction}
\label{section.01.introduction}

The  $\cW_N$ theories \cite{zam85,fz86b} are 2D conformal field theories (CFT) with  current algebra $\cW_N$, the representations of which are related to the weights of the Lie algebra $\sll_N$ (for a review, we refer the reader to \cite{bs95}). The Virasoro algebra  coincides with the $\cW_2$ algebra and is a sub-algebra of $\cW_{N>2}$. In the case of Virasoro CFT \cite{bpz84}, all correlation functions can be expressed in terms of correlation functions of primary fields. This is not true for the $\cW_{N>2}$ theory because the symmetry constraints are no longer sufficient to fix the operator product expansion or, equivalently, the matrix elements involving general states. In order to compute $\cW_N$ correlation functions one needs to find additional conditions besides the ones coming from the current algebra. Typically, these additional conditions can originate from the presence of a $\cW_N$ primary field, the null-state, at a certain level of a  given representation module. This is the reason why only  $\cW_N$  correlation functions  involving a set of particular fields can be computed.  For a general value of the central charge   a $\cW_N$ representation module can contain up to $N-1$ null-states: a field is said to be {\text fully-} or {\text semi-}  degenerate if the corresponding module contains $N-1$ or less null states. 

So far, the studies about $\cW_N$ theories focused mostly on $\cW_N$ correlation functions that contain level-1 semi-degenerate fields \cite{fl07c, fl08}. These fields are associated to the \textit{(anti-)fundamental} representation of $\slN$ and we will refer to them as (anti)-fundamental fields. 
In \cite{kms10} it was shown that all matrix elements that 
contain at least one (anti-)fundamental field can be computed explicitly. The interest about this special sector of $\cW_N$ was also greatly motivated by the fact that, via the AGT correspondence \cite{agt09,wyl09,mironov2010agt}, the corresponding conformal blocks are related to the instanton 
calculus in $\Omega$-deformed $\mathcal{N}=2$ SUSY $SU(N)$ quiver gauge theories.
On the symmetry level this relation is a consequence of the fact the $\cW_N$ (and many others) chiral symmetry algebras 
can be obtained from a special toroidal algebras having simple action on the cohomologies of instanton moduli spaces \cite{Belavin:2011pp}. This connection, in turn,  reveals a deep integrable structure of  $\cW_N$ theory \cite{aflt10,epss11} (for explicit example see, e.g. \cite{bb11}). In particular, the computation of the matrix elements of the semi-degenerate fundamental fields in the integrable basis \cite{aflt10} gives
a nice combinatorial representation for the conformal blocks.
Hence, an interesting problem is whether or not one can extend this connection to a larger set of  $\cW_N$ correlation functions.
The first question in this direction is whether there exists any generalization of the semi-degenerate fundamental field, such that the corresponding matrix elements can be also
constructed explicitly via $\cW_N$ symmetry constraints.

In a recent study \cite{BEFS16} we considered  a $\cW_3$ Toda 4-point functions that involve a fully-degenerate primary field in the fundamental representation of $\slthree$ and a fully-degenerate primary field in the \textit{adjoint} 
representation of $\slthree$. This latter field has {\it two} null-states at the second level of the associated representation module. In \cite{BEFS16}, we showed that the associated conformal block  satisfy a fourth-order Fuchsian differential equation and we discussed the role of multiplicities that appear in this theory. 
In this  paper  we generalize these result by considering 
the 4-point conformal block that involve, besides the fully-degenerate primary field in the fundamental representation, a semi-degenerate primary field with {\it one} null state at level two. The associated local correlation functions have been considered in \cite{fl08} where their expressions in term of four-dimensional integrals were given. Here we show that this conformal block obeys a sixth-order Fuchsian differential equation. We compute the matrix elements of the semi-degenerate field, between two arbitrary descendant states and we verify that the series expansion of the conformal block constructed from these matrix elements agrees with the differential equation. 

%SUBSECTION
%\subsection{Outline of contents}
The paper is organized as follows. In section 
\textbf{\ref{section.02.w3.toda.conformal.field.theory}}, 
we recall basic facts regarding $\cW_3$ conformal 
field theory and introduce the $\cW_3$ conformal  block function. In  \textbf{\ref{section.03.nullstates}} we discuss the null-vector conditions
for the degenerate and semi-degenerate primary fields.
In         \textbf{\ref{section.04.fuchsian.differential.equation}}, 
we focus on a specific 4-point correlation function with one semi-degenerate level=2 and one fully degenerate field in the fundamental  $\slthree$ representation. Here show that the corresponding conformal blocks obeys 
a sixth-order Fuchsian differential equation.   We proceed by discussing the matrix elements of the semi-degenerate level-2 field between 
two arbitrary descendant states and compare the results of the differential equation with the explicit construction of the conformal 
blocks in terms of the matrix elements.
In         \textbf{\ref{section.05.summary.comments}},
we present our conclusion and discuss some open problems.
In appendix   \textbf{\ref{detailderiv}},
we give some technical details of the derivation of the differential equation.

%SECTION.02
\section{$\cW_3$ chiral algebra, representation modules and conformal blocks }
\label{section.02.w3.toda.conformal.field.theory}

\label{w3.algebra}
We briefly introduce the $\cW_3$ chiral symmetry algebra and its representation theory. We use the same notations and normalization conventions as in \cite{BEFS16}.

\paragraph{$\cW_3$ algebra.}
The $\cW_3$ is an associative algebra generated by the modes $L_n$ and $W_n$ of the spin-$2$ 
energy-momentum tensor $\cT (z)$ and of the spin-$3$ holomorphic field $\cW(z)$. The full $\cW_3$ algebra 
is given by the following commutation relations

\begin{equation}
\left[ L_m, L_n \right]  =  (m-n) L_{m+n} + \frac{c}{12} (m^3-m) \delta_{m+n, 0}\;, \quad 
\left[ L_m, W_n \right] =  (2m-n) W_{m+n}\;,
\label{LW.commutator}
\end{equation}
and
\begin{multline}
\left[ W_m, W_n \right] = \frac13 (m-n) \Lambda_{m+n} 
\\
+
\ll \frac{22+5c}{48} \rr \ll \frac{m-n}{30}       \rr (2m^2-m n+2 n^2-8) \, L_{m+n}
\\
+
\ll \frac{22+5c}{48} \rr \ll \frac{c}{3 \cdot 5!} \rr (m^2-4) (m^3-m) \, \delta_{m+n, 0}\;,
\label{WW.commutator}
\end{multline}

\noindent where $\Lambda_m$ are the modes of the quasi-primary field 
$\Lambda = :\cT^2: - \frac{3}{10} \, \partial^2 \cT$ and
the colons $: \ :$ stand for normal-ordering. Explicitly, 

\begin{equation}
\Lambda_m = \sum_{p \leq -2} L_p  L_{m-p}  
          + \sum_{p \geq -1} L_{m-p} L_p  
          - \frac{3}{10} (m+2) (m+3) L_m\;.
\end{equation}

\noindent In \eqref{WW.commutator} we assume the following normalisation for the current $\cW(z)$ 
2-point function: 

\begin{equation}
\label{norm_w}
\langle \cW (1) \cW (0) \rangle = \frac{c}{3}\eta \quad \text{with} \quad \eta \mydef \ll \frac{22+5c}{48}\rr.
\end{equation}

The parametrisation of the $\cW_3$ central 
charge $c$, commonly used in the Toda field theory literature, is  

\begin{equation}
c = 2 + 24 \;Q^2\;, \quad 
Q = b + \frac{1}{b}\;.
\end{equation}

\paragraph{$\cW_3$ primary fields.}
A $\cW_3$ primary field $\Phi_{\vec{\alpha}}(z)$ is completely characterized by the pair of quantum numbers $(h,q)$, respectively its conformal dimension ($h$) and $W_0$ eigenvalue ($q$). It is labelled by a vector $\vec{\alpha}$ in the space spanned by the fundamental $\slthree$ weights,

\begin{equation}
\vec{\alpha} = \alpha_1 \; \vec{\omega}_1 + \alpha_2 \; \vec{\omega}_2\;,
\end{equation}
where the standard $\slthree$ conventions are used: 
\begin{align}
\vec{\omega}_1 ={}&  \sqrt{\frac23} \left(1, 0\right);\quad  \vec{\omega}_2 = \sqrt{\frac23} \left(\frac12, \sqrt{\frac34}\right);\\
\vec{e}_1 = {}& 2 \vec{\omega}_1 - \vec{\omega}_2;\quad 
\vec{e}_2 =  -\vec{\omega}_1 + 2 \vec{\omega}_2;
\quad \vec{\rho}  = \vec{\omega}_1 + \vec{\omega}_2,\\
\vec{h}_1 ={}&  \vec{\omega}_1;\quad 
\vec{h}_2 =  \vec{\omega}_1 - \vec{e}_1;\quad
\vec{h}_3 =  \vec{\omega}_1 - \vec{e}_1 - \vec{e}_2.
\end{align}

%\noindent  Noting as $\vec{P}$ as
%
%\begin{equation}
%\vec{\alpha} = \vec{P} + Q \vec{\rho}
%\end{equation} 
  
\noindent In terms of the parameters

\begin{equation}
x_i =  \left(Q\vec{\rho}-\vec{\alpha}\right) \cdot \vec{h}_i,  \quad i = 1, 2, 3, 
\end{equation}
one has

\begin{equation}
	h  = Q^2 + x_1 x_2 + x_1 x_3 + x_2 x_3\quad \text{and} \quad q = i \; x_1 x_2 x_3.
\end{equation}
The fields $\Phi_{\vec{\alpha}}$ are normalized in such a way that the 2-point correlation function $\langle  
\Phi^*_{\vec{\alpha}} (z) \Phi_{\vec{\alpha}} (0) \rangle $, the $\Phi^{*}_{\vec{\alpha}}(z)\mydef \Phi_{2Q \vec{\rho}- \vec{\alpha}}(z)$ satisfies:
\begin{equation}
\label{eq:defconj}
\lim_{z\to \infty} z^{2 h}\langle  
\Phi^*_{\vec{\alpha}} (z) \Phi_{\vec{\alpha}} (0) \rangle =1.
\end{equation}
In the following we will also use the notation $\vec{\alpha}^{*}\mydef 2Q \vec{\rho}- \vec{\alpha}$.

\paragraph{$\cW_3$ representation module.} The representation module $\mathcal{V}_{\vec{\alpha}}$ associated to $\Phi_{\vec{\alpha}}$ is spanned by the basis states,

\begin{equation}
\Phi^{(I)}_{\vec{\alpha}}\mydef \cL_I\;\Phi_{\vec{\alpha}} \mydef\; L_{-i_m} \cdots L_{-i_1} 
                            W_{-j_n} \cdots W_{-j_1} \Phi_{\vec{\alpha}} ,
\label{W3Verma}                            
\end{equation}

\noindent where the sets of positive integers
 
\begin{equation}
I=\{i_m, \cdots, i_1 ; j_n, \cdots, j_1\}\quad  \text{with}\quad i_m \geq \cdots \geq i_1 \geq 1, \quad 
j_n \geq \cdots \geq j_1 \geq 1 ,
\end{equation}
 are normal-ordered,
The symbol $\emptyset$ will be used when no modes $L_{m}$ or $W_n$ are present. For instance, set $I= \{\emptyset; 3,1\}$, then $\Phi^{(I)}_{\vec{\alpha}}= W_{-3}W_{-1}\Phi_{\vec{\alpha}}$.
%$\{i_n,\cdots,i_1; \emptyset\}$ when no $L_i$, and $W_i$ modes act on the state, respectively, 
%and $I = \{\emptyset, \emptyset\}$, when neither type of modes act on the state, 
%$\cL_{\emptyset} \; |h, q \rangle = |h, q\rangle$.  In the following we will use the notation 
%$\Phi^{(I)_X} =\cL_I \, \Phi_X$ to denote a descendant field associated to the primary field 
%$\Phi_{X}$, where $X$ indexes the quantum numbers $h,q$. For instance 
%$\Phi^{(\{2,1;3,1,1\})_1} =L_{-2}L_{-1}W_{-3}W_{-1}^2 \Phi_{h_1,q_1}$
\noindent The descendant fields $\Phi^{(I)}_{\vec{\alpha}}$ have conformal dimension $h+|I|$, where $|I|=i_1+i_2+\cdots+j_1+j_2+\cdots$ is called the level.  Any $\cW_3$ highest-weight representation is spanned by the states (\ref{W3Verma})\cite{bs95}. We also refer to the Appendix of \cite{BEFS16} where these properties are reviewed in the same notations and conventions adopted here. 

\paragraph{The Shapovalov matrix of inner products.}
The Shapovalov matrix $H$, whose $ij$-element $H_{ij}$ is the scalar product of the  states $|\cL_I\Phi_{\vec{\alpha}}\rangle$ and $|\cL_J\Phi_{\vec{\alpha}}\rangle$

\begin{equation}
\label{Shalapov}
H_{IJ}\ll \vec{\alpha}\rr = \langle \cL_I\Phi_{\vec{\alpha}} |\cL_J\Phi_{\vec{\alpha}} \rangle\;,
\end{equation}

\noindent has a block-diagonal structure, $H\ll \vec{\alpha}\rr=\text{diag} \ll H^{(0)}\ll \vec{\alpha}\rr, H^{(1)}\ll \vec{\alpha}\rr, H^{(2)}\ll \vec{\alpha}\rr, \cdots \rr$, 
where the elements of the $i$-th block, $H^{(i)}\ll \vec{\alpha}\rr$, are the scalar products of the level-$i$
descendants. These elements can be computed using the commutation relations
\eqref{WW.commutator},\eqref{LW.commutator}. By definition, $H^{(0)}\ll \vec{\alpha}\rr= 1$. The explicit forms of $H^{(1)}\ll \vec{\alpha}\rr$ and $H^{(2)}\ll \vec{\alpha}\rr$ can be found for instance in the  Appendix of \cite{BEFS16}.

\paragraph{The matrix elements.}
\label{kanno.algorithm}
The matrix elements of general
descendant fields are defined by:

\begin{eqnarray}
\label{gamma-gammap} 
\Gamma_{I, J, K} \ll L,M,R\rr
 &=& 
\frac{\langle 
(\Phi^{*}_{L})^{(I)}| 
\, 
\Phi^{(J)}_{M}(1) 
\, 
\Phi^{(K)}_{R}(0)
\rangle}{\langle 
(\Phi^{*}_{L})| 
\, 
\Phi_{M}(1) 
\, 
\Phi_{R}(0)
\rangle}\;,
\\
\Gamma^{\prime}_{I, J, K} \ll L,M,R\rr &=& 
\frac{\langle 
\Phi^{(I)}_{L}| 
\, 
\Phi^{(J)}_{M}(1) 
\, 
\Phi^{(K)}_{R}(0)
\rangle}{\langle 
\Phi_{L}| 
\, 
\Phi_{M}(1) 
\, 
\Phi_{R}(0)
\rangle} \; ,
\end{eqnarray}

\noindent where $L,M,R$ is a short notation for  $\vec{\alpha}_L, \vec{\alpha}_M, \vec{\alpha}_R$ and $\Phi^{(I)}_X$ ($X=L,M,R$) is defined in (\ref{W3Verma}).

\paragraph{Degenerate representations and fusion rules.} 
A $\cW_3$ fully-degenerate representation $\mathcal{V}_{r_1 r_2 s_1 s_2}$ is associated with the primary 
field $\Phi_{\vec{\alpha}_{r_1 r_2 s_1 s_2}}(z)$ with

\begin{equation}
\vec{\alpha}_{r_1 r_2 s_1 s_2} = 
b
\ll
(1 - r_1)\; \vec{\omega}_1 +
(1 - r_2)\; \vec{\omega}_2 
\rr 
+ 
\frac{1}{b}
\ll
(1 - s_1)\; \vec{\omega}_1 +
(1 - s_2)\; \vec{\omega}_2 
\rr,
\end{equation}

\noindent where $r_1, r_2, s_1,s_2$ are positive integers. In the following we use the notation:
\begin{equation}
\Phi_{r_1 r_2 s_1 s_2}\mydef\Phi_{\vec{\alpha}_{r_1 r_2 s_1 s_2}}.
\end{equation}
\noindent The representation $\mathcal{V}_{r_1 r_2 s_1 s_2}$ exhibits two independent null-states at levels $r_1 s_1$ and  $r_2 s_2$ and the fusion products of $\mathcal{V}_{r_1 r_2 s_1 s_2}$ with a general $\cW_3$ irreducible 
module $\Phi_{\vec{\alpha}}$ takes the form
\begin{equation}
\label{w3: fuprod}
\mathcal{V}_{r_1 r_2 s_1 s_2} \times \mathcal{V}_{\vec{\alpha}} =  
\sum_{\vec{h}_r, \vec{h}_s} \mathcal{V}_{\vec{\alpha} - b \vec{h}_r-b^{-1} \vec{h}_s},
\end{equation}

\noindent where $h_r$ and $h_s$ are the weights of the $\slthree$ representation with 
highest-weight \\$(r_1-1)\; \vec{\omega}_1 +
( r_2-2)\; \vec{\omega}_2$ and $(s_1-1)\; \vec{\omega}_1 +
( s_2-2)\; \vec{\omega}_2$ respectively.

\paragraph{$\cW_3$ conformal blocks.}
 
 The conformal block $\cB_M \ll L, 2, 1, R \rr (\{z_i\})$  with internal fusion channel $\Phi_{M}$ can be represented as the following \textit{comb diagram}:

\begin{equation}
\label{combdiag}
\begin{tikzpicture}
[line width=1.2pt]
\draw (-7.5,0)node[left]{$\cB_M \ll L, 2, 1, R \rr (\{z_i\})$};
\draw (-7.2,0)node[right]{$\mydef$};
\draw (-6.3,0)node[right]{$\langle \Phi^*_{L}(z_L)|\Phi_2 (z_2) \Phi_1 (z_1) \Phi_R(z_R) \rangle$};
\draw (-8.7,-3) node[right]{$\mydef$};
\draw (-6,-3)--(-5,-3);
\draw (-5,-3)--(-5,-2);
\draw (-5,-3)--(-3,-3);
\draw (-3,-3)--(-3,-2);
\draw (-3,-3)--(-2,-3);
\draw (-6,-3) node[left]{$\Phi^*_L (z_L)$};
\draw (-5,-2) node[above]{$\Phi_2 (z_2)$};
\draw (-3,-2) node[above]{$\Phi_1 (z_1)$};
\draw (-2,-3) node[right]{$\Phi_R (z_R)$};
\draw (-4,-3) node[below]{$M$};
\end{tikzpicture}
\end{equation}

\noindent Invariance under global conformal transformations implies:
\begin{eqnarray}
\cB_M \ll L, 2, 1, R \rr (\{z_i\}) &=&\left(z_L-z_1\right)^{-2  h_1} \left(z_L-z_R\right)^{h_1-h_R+h_2-h_L} \left(z_L-z_2\right)^{h_1+h_R-h_2-h_L} \times\nonumber \\
&\times& \left(z_2-z_R\right)^{-h_1-h_R-h_2+h_L}
 \cB_M \ll L, 2, 1, R \rr \left(z\right),
 \label{global_conformal}
\end{eqnarray}
where
\begin{equation}
\label{anharmonic_factor}
z =\frac{(z_1-z_R)(z_2-z_L)}{(z_1-z_L)(z_2-z_R)}\;.
\end{equation}
The  function $\cB_M \ll L, 2, 1, R \rr (z)$ is 
%a function of the holomorphic coordinate $z$ defined in (\ref{anharmonic_factor}) and it is 
defined by the following expansion

\begin{multline}
z^{h_L + h_R - h_M} \cB_M \ll L, 2, 1, R \rr (z)
=
\\
1 + 
\sum_{i=1}^\infty z^i 
\sum_{
\substack{
K, K' 
\\ 
|K| = |K'| = i
}
} 
 \left[H^{(i)}\ll M\rr\right]^{-1}_{K, K'} \Gamma_{\emptyset, \emptyset, K}\ll L,2,M\rr
 \Gamma^{\prime}_{K', \emptyset, \emptyset}\ll M,1,R\rr,
\label{cb.series.expansion}
\end{multline}

\noindent where the matrix $H^{(i)}$ and the matrix elements $\Gamma_{I_L,J_M,K_R}$ and $\Gamma^{\prime}_{I_L, J_M, K_R}$ were defined in (\ref{Shalapov}) and  (\ref{gamma-gammap}). 

\paragraph{$\cW_3$ Ward identities.}
While the three Ward identities associated with the conserved current $\mathcal{T}$ fix the form (\ref{global_conformal}),
there are other five Ward identities associated with the conserved current $\cW$
 
\begin{align}
\label{W: ward2_1}
&\sum_{X=L,2,1,R}           \, W_{-2}^{(X)}                                     \cB_M \ll L, 2, 1, R \rr (\{z_X\}) =0, \\
&\sum_{X=L,2,1,R} \ll z_{X}   \, W_{-2}^{(X)} +         \, W_{-1}^{(X)}             \rr\cB_M \ll L, 2, 1, R \rr (\{z_X\}) =0, \\
&\sum_{X=L,2,1,R} \ll z_{X}^2 \, W_{-2}^{(X)} + 2 z_X   \, W_{-1}^{(X)}+   q_{X} \rr \cB_M \ll L, 2, 1, R \rr (\{z_X\})=0,\\
&\sum_{X=L,2,1,R} \ll z_X^3 \, W_{-2}^{(X)} + 3 z_X^2 \, W_{-1}^{(X)}+3 z_X   q_X \rr\cB_M \ll L, 2, 1, R \rr (\{z_X\})=0, \\
\label{W: ward2_5}
&\sum_{X=L,2,1,R}^N \ll z_X^4 W_{-2}^{(X)} + 4 z_X^3 W_{-1}^{(X)}+6 z_X^2 q_X \rr \cB_M \ll L, 2, 1, R \rr (\{z_X\})=0,
\end{align}
where the notation $W_{-i}^{(X)}$, $i=1,2$ means that the mode $W_{-i}$ is applied to the  field $X(=L,2,1,R)$ in the conformal block.
%SUBSECTION

\section{Null-states  in semi- and fully-degenerate representations }
\label{section.03.nullstates}

%SUBSECTION
\subsection{Null-state equations for the fully-degenerate fundamental field}

We list here  the null-state equations for the fully-degenerate fundamental field $\Phi_{2111}(z)$  that we will need later. The field $\Phi_{2111}(z)=\Phi_{-b\vec{\omega}_1}(z)$ has quantum numbers

\begin{equation}
\label{hq2111}
h= h_1 \;\mydef \;
\frac13 \ll - 3 - 4 b^2 \rr \;,
\quad 
q = q_1 \;\mydef\;
\frac{i}{27 b} \ll 3+4b^2 \rr \ll 3 + 5 b^2 \rr \;.
\end{equation}

\noindent This field obeys level-1, level-2 and level-3 null-state conditions: 

\begin{eqnarray}
W_{-1} \Phi_{-b\vec{\omega}_1} &=& \frac{3 q_1}{2h_1} L_{-1} \Phi_{-b \vec{\omega}_1}.
\label{Level.1.identity}
\\ 
W_{-2} \Phi_{-b \vec{\omega}_1}  &=& 
\ll \frac{12 q_1             }{h_1 (5 h_1 + 1)} L_{-1}^2 - 
    \frac{ 6 q_1 (h_1+1)}{h_1 (5 h_1 + 1)} L_{-2} \rr   \Phi_{-b \vec{\omega}_1} ,
\label{Level.2.identity}
\end{eqnarray}

\noindent and

\begin{multline}
\label{Level.3.identity}
W_{-3}\Phi_{-b \vec{\omega}_1}  =
\ll\frac{16 q_1}{ h_1 (h_1+ 1) (5 h_1 +1)} L_{-1}^3
\right.
\\
\left.
- \, \frac{12 q_1}{h_1 (5 h_1 + 1)} L_{-1} L_{-2}+\frac{3q_1(h_1-3)}{2 h_1 (5 h_1 + 1)}
L_{-3} \rr\Phi_{-b\vec{\omega}_1}.
\end{multline}

\subsection{Semi-degenerate  representation at level-two}
\label{semidegnull.states}

In this paper we are interested in the case of general central charge, in particular when $b^2 \notin \mathbb{Q}$. Consider a  field $\Phi_{r_1 r_2 s_1 s_2}$ with $r_1$ and $s_1$ positive integers and $r_2 \notin \mathbb{N}^{+}$. In the representation module there is one null-vector at level $r_1 s_1$ with quantum numbers $(h,q)$ coinciding with the ones of the primary operator $\Phi_{r'_1 r'_2 s_1 s_2}$ with $r_1'=-r_1$ and $r'_2= r_2+r_1$, see \cite{bs92} and references therein. Here we want to focus on the semi-degenerate field at level two $\Phi_{-b \vec{\omega}_1+ s \vec{\omega}_2}$, which corresponds to $\Phi_{r_1 r_2 s_1 s_2}$ with $r_1 =2$ and $r_2 = 1-s$. In \cite{bw92}, the explicit expression of the null-vector at level $r$ associated to the field $\Phi_{(1-r)b \vec{\omega}_1+s s \vec{\omega}_2}$ has been given. For sake of clarity, we derive below, by using our conventions, the same result for the level two null-vector associated to the field $\Phi_{-b \omega_1+ s \omega_2}$.

Let consider a general module $\mathcal{V}_{\vec{\alpha}}$ with quantum number $(h,q)$. At the second level one has five fields $\Phi^{(I)}$ with $|I|=2$:
\begin{equation}
|I|=2:\quad  I=\{2;\emptyset\}, \{\emptyset; 2\}, \{ 1,1;\emptyset\},\{ 1;1\}, \{\emptyset; 1,1\}.
\end{equation}
From (\ref{W3Verma}), a general field $\Psi$ at level two can be written as a linear combination of the above five fields:
\begin{equation}
\Psi_{\vec{\alpha}} \mydef 
\ll 
c_{\{2;\emptyset\}}  L_{-2} +
c_{\{\emptyset;2\}}  W_{-2}   +
c_{\{1,1;\emptyset\}}  L_{-1}^2 + 
c_{\{1;1\}}  L_{-1} W_{-1} + W_{-1}^2 
\rr 
\Phi_{\vec{\alpha}} ,
\end{equation} 
where the global normalisation has been fixed by setting the coefficient of $W_{-1}^2\Phi_{\vec{\alpha}}$ to one.
\noindent We assume  $\Psi$ to be a $\cW_3$ primary field, i.e.
\begin{eqnarray}
W_0 \Psi_{\vec{\alpha}} &=& q' \Psi_{\vec{\alpha}}\;, \label{Q0cond}  \\
L_{1}\Psi_{\vec{\alpha}} &=& L_{2}\Psi_{\vec{\alpha}}=0\;. \label{L1L2cond}
\end{eqnarray} 
We will see below that the above relations, in particular the fact that $W_0$ is diagonalizable, give the correct result. A discussion about the fact that $W_{0}$ need not be diagonalizable can be found for instance in \cite{wattsdet}. 
The  annihilation by  $W_1$ and $W_2$ and higher generators follows from the $\cW_3$ commutation
relations. 

\noindent The requirement (\ref{Q0cond}) is satisfied if
\be\label{Q0eigenvec}
M u=q' u\;,
\ee
 where $c$ is the vector 
$u=(u_{\{2;\emptyset\}}, u_{\{\emptyset; 2\}}, u_{\{ 1,1;\emptyset\}},u_{\{ 1;1\}}, u_{\{ \emptyset;1,1\}})$ and the matrix $M$ reads
\begin{equation}
M=\ll 
\begin{array}{ccccccccc}
q & & \frac{4h}{3}  & & 0 & & 0                               & & 2q\\
4 & & q                    & & 2 & & 0                               & & \frac{2-c+32h}{48}\\
0 & & \frac{2}{3}    & & q & & \frac{2-c+32h}{48}  & & 0 \\
0 & & 0                    & & 4 & & q                               & & \frac{18-c+32h}{24}\\
0 & & 0                    & & 0 & & 2                               & & q 
\end{array} 
\rr\;.
\end{equation}
The requirement (\ref{L1L2cond}) is equivalent to the condition 
\be\label{nullcond}
N u=0\;,
\ee
where
\begin{equation}
N=\ll
\begin{array}{ccccc}
4h+\frac{c}{2}& 6q & 6 h & 9 q & \frac{5h(2-c+32h)}{48}\\
3 & 0  & 2(2 h+1) & 3q       &\frac{2-c+32h}{16}  \\
0 & 4  & 0        &  2 (h+1) & 6q \\
\end{array} 
\rr\;.
\end{equation}
We consider now the field $\Phi_{-b \vec{\omega}_1+ s\vec{\omega}_2}$ that has quantum numbers
\begin{equation}
\label{quantumnumberssemi}
h = h_2\;\mydef\; \frac{3 (s - b) + b(s-2 b )^2}{3 b},\quad q=q_2\;\mydef \;i\frac{(b+s) \left(4 b^2-2 b s+3\right) \left(5 b^2-b s+3\right)}{27 b^2}.
\end{equation}

\noindent We have found that the vector 
\begin{multline}
\label{csing}
u^{\text{sing}}=\left(-\frac{ (1 + 2 b^2 - b s) (2 + 2 b^2 - b s)}{3}, -\frac{ i (1 + 3 b^2) (3 + 4 b^2 - 2 b s)}{6 b},\right. \\
\left.-\frac{(3 + b^2 - 2 b s) (3 + 7 b^2 - 2 b s)}{36 b^2}, -\frac{ i (3 + 4 b^2 - 2 b s)}{3 b}, 1\right)
\end{multline}
satisfies the conditions \eqref{Q0eigenvec} and \eqref{nullcond} with the following eigenvalue
\be
q'_{\text{sing}}=\frac{i (5 b - s) (3 + 4 b^2 - 2 b s) (-3 + b^2 + b s)}{27 b^2}\;.
\ee
One can directly verify that $q'_{\text{sing}}$ is indeed the $W_{0}$ eigenvalue of the field $\Phi_{3 b \omega_1 + (-2 + s) b \omega_2}$, see the discussion at the beginning of the section.   
We have therefore shown that the field $\Phi_{-b \vec{\omega}_1+ s\vec{\omega}_2}$ has  a singular-state at level two. 
\noindent Decoupling of the singular vector $\Psi_{-b\vec{\omega}_1+s\vec{\omega}_2}=0$ yields the  following null-state condition
\begin{equation}
\label{nullvector2}
\sum_{\substack{
I
\\ 
|I| = 2}} u^{\text{sing}}_{I} \quad \Phi^{(I)}_{-b \vec{\omega}_1+ s\vec{\omega}_2}=0\;.
\end{equation}
In a similar way, a similar null-vector condition for the other semi-degenerate field at level two, $\Phi_{s \vec{\omega}_1-b \vec{\omega}_2}$ can be obtained. The remaining two other cases are obtained by replacing $b\to 1/b$.

%SUBSECTION

\section{Sixth-order differential equation}
\label{section.04.fuchsian.differential.equation}

We consider here the conformal block defined in (\ref{combdiag}) with the following identifications
\begin{equation}
\Phi_{2}=\Phi_{-b \vec{\omega}_1+s\vec{\omega}_2}, \quad \Phi_{1}=\Phi_{-b\vec{\omega}_1}\;.
\end{equation}
\noindent We use a shorter notation for the conformal block,  $\cB_M \ll L, 2, 1, R \rr (z)\to \cB_M(z)$:
\begin{equation}
\label{cbconsidered}
\begin{tikzpicture}
[line width=1.2pt]
\draw (-2.5,0)node[left]{$\cB_M (z)$};
\draw (-2.5,0) node[right]{$\mydef$};
\draw (0,0)--(1,0);
\draw (1,0)--(1,1);
\draw (1,0)--(3,0);
\draw (3,0)--(3,1);
\draw (3,0)--(4,0);
\draw (0,0) node[left]{$\Phi^*_L (\infty)$};
\draw (1,1) node[above]{$\Phi_{-b \vec{\omega}_1+s\vec{\omega}_2} (1)$};
\draw (3,1) node[above]{$\Phi_{2111} (z)$};
\draw (4,0) node[right]{$\Phi_R (0)$};
\draw (2,0) node[below]{$M$};
\end{tikzpicture}
\end{equation}
\noindent In the following we present in full detail the computation of this conformal block. The case in which $\Phi_2= \Phi_{s \vec{\omega}_1-b \vec{\omega}_2}$ will be briefly discussed in section 5.2.3.

\noindent Let us sketch the procedure  to obtain the differential equation satisfied by the conformal block defined above. At the beginning we have 
\begin{equation}
9\quad \text{unknown functions}:\quad \cB_M (z)\quad \text{and} \quad W_{-i}^{(X)}\cB_M (z) \quad i=1,2, X=L,2,1,R.
\nonumber 
\end{equation} We recall that the function $W_{-i}^{(X)}\cB_M (z)$ is the conformal block that involves the descendant $W_{-i} \Phi_{X}$ (for some explicit examples see the Appendix \ref{detailderiv}). 
\noindent Using the 8 equations coming from the 5 Ward identities (\ref{W: ward2_1})-(\ref{W: ward2_5}) plus the 3 null-state conditions (\ref{Level.1.identity}), (\ref{Level.2.identity}) and (\ref{Level.3.identity}), we arrive to an equation of the form
\begin{equation}
\label{1.equation}
\left[\partial_{z}^3+\cdots\right] \cB_M (z)=\cdots W_{-1}^{(2)}\cB_M (z)\;,
\end{equation}
where on the LHS the dots stand for a certain linear combination of differential operators,  for instance  $z^{-1}\partial_{z}^2$, $(z-1)^{-2}\partial_{z}^2$, $(z-1)^2\partial_{z}, (z-1)^{-2}$ etc., acting on $\cB_M  (z)$. On the RHS the dots also represent some known linear combination of the factors  $z^{-3}$, $(z-1)^{-3}$, $z^{-2}(z-1)^{-1}$ and $z^{-1}(z-1)^{-2}$ multiplying  $W_{-1}^{(2)}\cB_M (z)$. 
We notice that if we had identified  $\Phi_{2}=\Phi_{s \vec{\omega}_1}$, which is semi-degenerate at level 1,  we could have used the  relation $W_{-1} \Phi_{s \vec{\omega}_1}(z)\propto \partial_{z}\Phi_{s \vec{\omega}_1}(z)$ in the  eq.(\ref{1.equation}) and obtained  the third-order generalized hypergeometric differential equation of \cite{fl07c}. For general $\cW_N$ conformal blocks,  containing one fully-degenerate and one semi-degenerate fundamental field, one obtains a $N$-order generalized hypergeometric equation \cite{fl07c}. A detailed study of the $\cW_4$ theory can be found  in \cite{furlan2015some}. In  the case under consideration here,  eq.(\ref{cbconsidered}), the null-vector condition (\ref{nullvector2}) relates states at level $2$. 
We repeat  the procedure leading to (\ref{1.equation}), this time for the fields: 
\begin{equation}
W_{-1}^{(2)}\cB_M (z) \quad \text{plus} \quad 8 \quad \text{unknown functions}: \quad W_{-i}^{(X)}W_{-1}^{(2)}\cB_M (z) \quad i=1,2, X=L,2,1,R.
\nonumber
\end{equation}
\noindent Notice that in total we have 17 unknown functions. Again we use the 8 equations expressing the three null-state conditions (\ref{Level.1.identity}), (\ref{Level.2.identity}) and (\ref{Level.3.identity}) and the five Ward identities applied to the function $W_{-1}^{(2)}\cB_M(z)$. With respect to the eqs. (\ref{W: ward2_1})-(\ref{W: ward2_5}), these five Ward identities have additional terms that originate from the fact that $W_{-1}^{(2)}\cB_M(z)$ involves a descendant state. Restoring the dependence on the coordinates $z_L, z_2,z_1,z_R$, the modified Ward identities take the form

\begin{align}
\label{WW: ward2_1}
&\sum_{X=L,2,1,R}           \, W_{-2}^{(X)}                                     W_{-1}^{(2)}\cB_M (\{z_X\}) =0, \\
&\sum_{X=L,2,1,R} \ll z_{X}   \, W_{-2}^{(X)} +         \, W_{-1}^{(X)}             \rr W_{-1}^{(2)}\cB_M  (\{z_X\}) =0, \\
&\sum_{X=L,2,1,R} \ll z_{X}^2 \, W_{-2}^{(X)} + 2 z_X   \, W_{-1}^{(X)}+   q_{X} \rr W_{-1}^{(2)}\cB_M (\{z_X\})+\nonumber \\
&+\kappa \;\partial_{z_2}\cB_M (\{z_X\})=0,\\
&\sum_{X=L,2,1,R} \ll z_X^3 \, W_{-2}^{(X)} + 3 z_X^2 \, W_{-1}^{(X)}+3 z_X   q_X \rr W_{-1}^{(2)}\cB_M (\{z_X\})+\nonumber \\
&+ 3 \;z_2\; \kappa\; \partial_{z_2}\cB_M (\{z_X\}) + h_{2}\; \kappa \cB_M (\{z_X\})=0 , \\
\label{WW: ward2_5}
&\sum_{X=L,2,1,R}^N \ll z_X^4 W_{-2}^{(X)} + 4 z_X^3 W_{-1}^{(X)}+6 z_X^2 q_X \rr W_{-1}^{(2)}\cB_M (\{z_X\}) +\nonumber \\
&+ 6\; z_2^2\; \kappa\; \partial_{z_2}\cB_M (\{z_X\}) + 4\; h_{2}\; \kappa\; \cB_M (\{z_X\})=0\;,
\end{align}
where
\begin{equation}
\kappa = \frac{1}{48} \left(2 - c + h_2\right),
\end{equation} 
and the dimension $h_2$ is given in (\ref{quantumnumberssemi}).
This time we obtain the relation of the type
\begin{equation}
\label{2.equation}
\left[\partial_{z}^3+\cdots\right] W_{-1}^{(2)}\cB_M (z)=\cdots \left[W_{-1}^{2}\right]^{(2)}\cB_M(z)\;.
\end{equation}
Note that the above manipulations are valid for a general field $\Phi_{2}(x)$.  We can now use the fact that $\Phi_{2}=\Phi_{-b \vec{\omega}_1+s\vec{\omega}_2}$ and use (\ref{nullvector2}) in order to express $\left[W_{-1}^{2}\right]^{(2)}\cB_M(z)$ in terms of the functions  $\cB_M(z)$ and $W_{-1}^{(2)}\cB_M(z)$
\begin{equation}
\label{3.equation}
\left[W_{-1}^{2}\right]^{(2)}\cB_M(z) = \cdots  \cB_M^{''}(z)+\cdots z^{-1}\cB_M'(z)+\cdots+ \cdots \partial_{z} W_{-1}^{(2)}\cB_M(z)+\cdots z^{-1}W_{-1}^{(2)}\cB_M(z)+\cdots
\end{equation}

Using the system of equations (\ref{1.equation}), (\ref{2.equation}) and (\ref{3.equation}), we finally obtained  a sixth-order differential equation for the function $\cB_{M}(z)$. More details of the calculations are collected in the Appendix \ref{detailderiv}. 
However, the final and explicit expression of the sixth-order differential is too long to be reported here. We discuss instead the results that follow from it.

\section{Local exponents and matrix elements}
Defining 

\begin{equation}
\vec{\alpha}_R = a_{R_1} \vec{\omega}_1 + a_{R_2} \vec{\omega}_2,\quad \vec{\alpha}_L = a_{L_1} \vec{\omega}_1 + a_{L_2} \vec{\omega}_2\;,
\end{equation}

\noindent the conformal block $\cB_M (z)$ is a function of five parameters, $a_{R_1}$,
$a_{R_2}$, $a_{L_1}$,
$a_{L_2}$ and $b$.

\subsection{Local exponents from the differential equation}
The Fuschsian differential equation of order six has $2+1$ singularities at $0, 1$ and $\infty$. We refer the reader to \cite{yoshida1987fuchsian} for an exhaustive overview of 
Fuchsian systems. In Riemann-symbol notation
the local exponents $\rho^{0}_i,\rho^{1}_i$ and $\rho^{\infty}$, $i=1,\cdots,6$, associated to the $2+1$ singular points $0,1$ and $\infty$ can be represented as

\begin{equation}
\label{Riem_not}
\begin{Bmatrix}  
0& & 1 & & \infty  \\ 
\alpha_1   & &  \beta_1   & & \gamma_1   \\
\alpha_1 +1  & &  \beta_1 +1   & & \gamma_1 +1   \\
\alpha_2   & &  \beta_1+2  & & \gamma_2   \\
\alpha_2 +1 & &  \beta_2 & & \gamma_2 +1  \\
\alpha_3 && \beta_3 && \gamma_3 \\
\alpha_3 +1 && \beta_3 +1  && \gamma_3 +1
\end{Bmatrix}
\end{equation}

\noindent with

\begin{eqnarray}
\label{alphas}
\alpha_1 &=&\frac13 \left(2 a_{R_1} b + a_{R_2} b\right),
\quad
\alpha_2 =\frac13 \left(3 - a_{R_1} b + a_{R_2} b + 3 b^2\right),  
\quad   
\alpha_3 =\frac13 \left(6 - a_{R_1} b - 2 a_{R_2} b + 6 b^2\right) ,  
\nonumber \\
\label{betas}
\beta_1 &=& \frac13 \left(-2 b^2 + b s\right),
\quad \beta_2 =\frac13 \left(3 + 4 b^2 + b s\right), 
\quad
\beta_3 = \frac13 \left(6 + 7 b^2 - 2 b s\right)  ,\nonumber \\
\label{gammas}
\gamma_1 &=& \frac13 \left(a_{L_1}b - a_{L_2}b - 5 b^2-3\right),\,
\gamma_2 = \frac13 \left(-6 + a_{L_1} b + 2 a_{L_2} b - 8 b^2\right),\,
\gamma_3 = \frac13 \left(-2 a_{L_1} b - a_{L_2} b - 2 b^2\right) .  \nonumber \\
&&
\end{eqnarray} 
It is easily checked that the above local exponents satisfy the Fuchs identity:
\begin{equation}
\sum_{i=1}^{n} \left(\rho^0_{i}+\rho^1_{i}+\rho^\infty_{i} \right) = (k-1)\frac{n(n-1)}{2} =15,
\end{equation}
specified in our case where Fuchsian equation has order $n=6$ and  number of singularity  $2 (=k) +1$. 
It is important to observe that in our Fuchsian equation the fact that there are local exponent different by integers do not imply logarithmic solutions. Indeed, we argue below that the structure of local exponents does not origin from logarithmic features of the theory but from the undetermined matrix elements in the $\cW_3$ algebra.
\noindent We can verify that the values of the local exponents  can be found by using the fusion products (\ref{w3: fuprod}).
To this end we note that the solutions of the differential equation with diagonal monodromies  around $0,1,\infty$  define
correspondingly $s,t,u$ conformal blocks associated with the following diagrams:

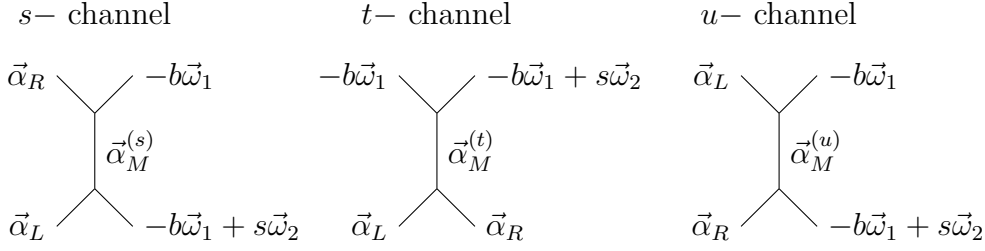
\begin{tikzpicture}[scale=0.5]
[line width=1.2pt]
\draw (1,1) node[above]{$s-$ channel};
\draw (0,0) node[left]{$\vec{\alpha}_{R}$};
\draw (0,0)--(1,-1);
\draw (2,0) node[right]{$-b \vec{\omega}_1$};
\draw (2,0)--(1,-1);
\draw (1,-2) node[right]{$\vec{\alpha}_{M}^{(s)}$};
\draw (1,-1)--(1,-3);
\draw (0,-4)--(1,-3);
\draw (0,-4) node[left]{$\vec{\alpha}_L $};
\draw (2,-4)--(1,-3);
\draw (2,-4) node[right]{$- b\vec{\omega}_1 + s\vec{\omega}_2 $};

\draw (10,1) node[above]{$t-$ channel};
\draw (9,0) node[left]{$-b \vec{\omega}_1$};
\draw (9,0)--(10,-1);
\draw (10,-2) node[right]{$\vec{\alpha}_{M}^{(t)}$};
\draw (11,0) node[right]{$- b\vec{\omega}_1 + s\vec{\omega}_2$};
\draw (11,0)--(10,-1);
\draw (10,-1)--(10,-3);
\draw (9,-4)--(10,-3);
\draw (9,-4) node[left]{$\vec{\alpha}_L $};
\draw (11,-4)--(10,-3);
\draw (11,-4) node[right]{$\vec{\alpha}_R  $};

\draw (19,1) node[above]{$u-$ channel};
\draw (18,0) node[left]{$\vec{\alpha}_{L}$};
\draw (18,0)--(19,-1);
\draw (19,-2) node[right]{$\vec{\alpha}_{M}^{(u)}$};
\draw (20,0) node[right]{$-b \vec{\omega}_1$};
\draw (20,0)--(19,-1);
\draw (19,-1)--(19,-3);
\draw (18,-4)--(19,-3);
\draw (18,-4) node[left]{$\vec{\alpha}_R $};
\draw (20,-4)--(19,-3);
\draw (20,-4) node[right]{$- b\vec{\omega}_1 + s\vec{\omega}_2 $};

\end{tikzpicture}

\noindent and that we denote by $\cB_M^{(s)}(z)$, $\cB_M^{(t)}(z)$ and $\cB_M^{(u)}(z)$. The crossing symmetry relation relates 
\be
\cB_M^{(s)}(z)\leftrightarrow\cB_M^{(t)}(1-z)\leftrightarrow z^{-2 h_1}\cB_M^{(u)}(1/z)\;.
\ee
Note, that in order to reconstruct $s,t,u$ fusion rules from the differential equation, one has to take into account
factors coming form the fields transformation (non trivial only in $z\rightarrow1/z$ transformation) . Substituting one of these
there functions in the differential equation and keeping leading terms in the expansions around $0,1,\infty$ we find that
the exponents $\alpha_i$, $\beta_i$ and $\gamma_i$, associated respectively to the $s-$, $t-$, and $u-$ channels, correspond to the following fusion channels:
\begin{itemize}
\item  $s-$channel:

\begin{eqnarray}
\label{ch1}
\text{Channel 1,2}\;\; (\alpha_1):        \quad 
\vec{\alpha}_M^{(s)} &=& \vec{\alpha}_R- b \,     \vec{\omega}_1 \\
\label{ch2}
\text{Channel 3,4        }\; (\alpha_2):\quad 
      \vec{\alpha}_M^{(s)} &=& \vec{\alpha}_R+ b \, \ll \vec{\omega}_1 - \vec{\omega}_2 \rr   \\
\label{ch34}
\text{Channel 5,6        }\; (\alpha_3):\quad
\vec{\alpha}_M^{(s)} &=& \vec{\alpha}_R+ b \,     \vec{\omega}_2  
\end{eqnarray}

\item $t-$channel:
\begin{eqnarray} 
\label{ch1t}
\text{Channel 1,2,3}\; (\beta_1):        \quad 
\vec{\alpha}_M^{(t)} &=& \ll - b\; \vec{\omega}_1+s\;\vec{\omega}_2\rr \,-b     \vec{\omega}_1 \\
\label{ch2t}
\text{Channel  4}\;\; (\beta_2):\quad 
      \vec{\alpha}_M^{(t)} &=& \ll - b\; \vec{\omega}_1+s\;\vec{\omega}_2\rr+ b \, \ll \vec{\omega}_1 - \vec{\omega}_2 \rr     \\
\label{ch34t}
\text{Channel 5,6        }\; (\beta_3):\quad
\vec{\alpha}_M^{(t)} &=&\ll - b\; \vec{\omega}_1+s\;\vec{\omega}_2\rr+ b \,     \vec{\omega}_2  
\end{eqnarray}
\item   $u-$channel:

\begin{eqnarray}
\label{ch1u}
\text{Channel 1,2}\;\; (\gamma_1):        \quad 
\vec{\alpha}_M^{(u)} &=& \vec{\alpha}_L^*- b \,     \vec{\omega}_1 \\
\label{ch2u}
\text{Channel  3,4        }\; (\gamma_2):\quad 
      \vec{\alpha}_M^{(u)} &=& \vec{\alpha}_L^*+ b \, \ll \vec{\omega}_1 - \vec{\omega}_2 \rr   \\
\label{ch34u}
\text{Channel 5,6       }\; (\gamma_3):\quad
\vec{\alpha}_M^{(u)} &=& \vec{\alpha}_L^*+ b \,     \vec{\omega}_2  
\end{eqnarray}

\end{itemize}
We note that as usual (in  $\cW_3$ case) in the $u$-channel we use the conjugated  value $\vec{\alpha}_L^*$ which is defined below  (\ref{eq:defconj}).

\subsection{Multiplicities and matrix elements}

The multiplicities of the local exponents can be argued from the 4-point conformal block expansion (\ref{cb.series.expansion}).
As explained in \cite{fl07c,bw92,bpt92,bw93, wat94, kms10,BEFS16}  any matrix element $\Gamma_{I,J,K}$ of three arbitrary $\cW_3$ 
descendant states, $\Phi^{(I)}_L$, $\Phi^{(J)}_M$, and $\Phi^{(K)}_R$ can be written as linear combinations of the matrix elements 
\begin{equation}
\label{me_basis}
\Gamma_{\{\emptyset;\emptyset\},\{\emptyset;\underbrace{1,1,1,\cdots}_{p \; \text{times}}\},\{\emptyset;\emptyset\} }\ll \vec{\alpha}_{L}, \vec{\alpha}_M, \vec{\alpha}_{R}\rr, \quad  p=1,2,3,\cdots
\end{equation} 
where the $\Gamma$ have been defined in (\ref{gamma-gammap}). 

\subsubsection{Multiplities in the $s-$ and $u-$channel}

Consider for instance the conformal block expansion (\ref{cb.series.expansion}) in the $s-$channel. To determine the first order coefficient one needs to evaluate the two matrix elements:
\begin{equation}
\Gamma_{\{\emptyset;\emptyset\},\{\emptyset;1\},\{\emptyset;\emptyset\} }\ll \vec{\alpha}_{M}^{(s)}, -b\vec{\omega}_1, \vec{\alpha}_{R}\rr,\quad \Gamma'_{\{\emptyset;\emptyset\},\{\emptyset;1\},\{\emptyset;\emptyset\} }\ll \vec{\alpha}_{L}, -b\vec{\omega}_1+s \vec{\omega}_2, \vec{\alpha}^{(s)}_{M}\rr.
\end{equation}
If the $\Gamma$ matrix element, involving the fully-degenerate field $\Phi_{-b \vec{\omega}_{1}}$ can be evaluated using the null-state equation (\ref{Level.1.identity}):
\begin{equation}
\Gamma_{\{\emptyset;\emptyset\},\{\emptyset;1\},\{\emptyset;\emptyset\} }\ll \vec{\alpha}_{M}^{(s)}, -b\vec{\omega}_1, \vec{\alpha}_{R}\rr = \frac{3 q_{1}}{2 h_{1}}\left(q^{(s)}_{M}-q_R-q_{1}\right),
\end{equation}
the matrix element $\Gamma'_{\{\emptyset;\emptyset\},\{\emptyset;1\},\{\emptyset;\emptyset\} }\ll \vec{\alpha}_{L}, -b\vec{\omega}_1+s \vec{\omega}_2, \vec{\alpha}^{(s)}_{M}\rr$ is determined only if $s=-b$ (and $h_{L}\neq h^{(s)}_{M}$), i.e. when one of the fields entering the matrix elements is a fully-degenerate field in the adjoint representation. This case was fully considered in \cite{BEFS16}. For $s\neq -b$ instead, the condition at second level (\ref{nullvector2}) alone is not sufficient to fix this matrix element. This ambiguity is at the origin of the multiplicity of order two of the local exponents $\alpha_{i}$. Indeed, let us fix 
\begin{equation}
\label{mep1}
\Gamma'_{\{\emptyset;\emptyset\},\{\emptyset;1\},\{\emptyset;\emptyset\} }\ll \vec{\alpha}_{L}, -b\vec{\omega}_1+s \vec{\omega}_2, \vec{\alpha}^{(s)}_{M}\rr= \lambda,
\end{equation} 
where $\lambda$ is an arbitrary constant. This is equivalent, from the point of view of the differential equation, to choose a particular combination in the two-dimensional space of solutions having the same local exponent $\alpha_i$. On the other hand, the semi-degenerate condition (\ref{nullvector2}) allows to express all the other matrix elements (\ref{me_basis}) with $p=2,3..$ as functions of (\ref{mep1}). For instance,
\begin{align}
\label{p2p1}
&\Gamma'_{\{\emptyset;\emptyset\},\{\emptyset;1,1\},\{\emptyset;\emptyset\} }\ll \vec{\alpha}_{L} , -b\vec{\omega}_1+s\vec{\omega}_2, \vec{\alpha}_{M}^{(s)}\rr = -c^{\text{sing}}_{\{2;\emptyset\}}  
\left( -h_L+2h^{(s)}_{M}+h_2\right)+\nonumber \\&+ c^{\text{sing}}_{\{\emptyset;2\}}\left(q_{M}^{(s)}+q_L+q_2+2\lambda\right)-c^{\text{sing}}_{\{1,1;\emptyset\}}\left((h_L-h_M^{(s)}-h_2-1)(h_L-h_M^{(s)}-h_2) \right)-\nonumber \\
&-c^{\text{sing}}_{\{1;1\}}\left((h_L-h_M^{(s)}-h_2-1)\lambda\right)\;.
\end{align}
We have checked up to the second order in the expansion (\ref{cb.series.expansion}), that the direct computation of the matrix element is in agreement with the results obtained by using the sixth-order differential equation.  The same arguments seen for explaining the multiplicities in the $s-$channel holds for the $u-$channel.

\subsubsection{Multiplicities in the $t-$channel}

One can notice that in the $t-$channel the structure of the multiplicities is different. Again this can understood by considering the matrix elements entering in the $t-$channel expansions.  For the fusion channels $1,2,3$, we have two undetermined matrix elements  at first and second order:
\begin{equation}
\Gamma'_{\{\emptyset;\emptyset\},\{\emptyset;1\},\{\emptyset;\emptyset\} }\ll \vec{\alpha}_{L} , -2 b\vec{\omega}_1+s\vec{\omega}_2, \vec{\alpha}_{R} \rr, \quad \Gamma'_{\{\emptyset;\emptyset\},\{\emptyset;1,1\},\{\emptyset;\emptyset\} }\ll \vec{\alpha}_{L} , -2 b\vec{\omega}_1+s\vec{\omega}_2, \vec{\alpha}_{R} \rr.
\end{equation} 
At the level of the differential equation, this corresponds to the fact that, in order to select a function in the three-dimensional space of solutions with local exponent $\beta_1$, we have to fix  two parameters. Fixing these two parameters is equivalent to fixing the values of the above matrix elements. Once these two parameters have been set, all the other matrix elements of higher-order can be computed in terms of these two parameters, as we have seen before. This is consistent with the fact that the field $\Phi_{-2 b\vec{\omega}_1+s\vec{\omega}_2}$ obeys a null-state condition at order three, see the discussion at the beginning of section 3.2.  In this respect, the differential equation is the most direct method to determine these matrix elements. 
Finally, the fact that the space of solutions with local exponent $\beta_2$ is uni-dimensional is due to the fact that all the matrix elements, even those at the first level, are known. Indeed in this case $\vec{\alpha}_{M}^{(t)} = (s-b)\vec{\omega}_2$. It is a semi-degenerate anti-fundamental field and the corresponding matrix elements (\ref{me_basis})  involve this field can be evaluated for any $p$. 

\subsubsection{The case with $\Phi_{s \vec{\omega}_1- b \vec{\omega}_2}$}
We consider now the conformal block:
\begin{equation}
\label{cbconsidered2}
\begin{tikzpicture}
[line width=1.2pt]
\draw (-2.5,0)node[left]{$\cB_M (z)$};
\draw (-2.5,0) node[right]{$\mydef$};
\draw (0,0)--(1,0);
\draw (1,0)--(1,1);
\draw (1,0)--(3,0);
\draw (3,0)--(3,1);
\draw (3,0)--(4,0);
\draw (0,0) node[left]{$\Phi^*_L (\infty)$};
\draw (1,1) node[above]{$\Phi_{s \vec{\omega}_1-b\vec{\omega}_2} (1)$};
\draw (3,1) node[above]{$\Phi_{2111} (z)$};
\draw (4,0) node[right]{$\Phi_R (0)$};
\draw (2,0) node[below]{$M$};
\end{tikzpicture}
\end{equation}
where we consider the other semi-degenerate field at level two, $\Phi_{s \vec{\omega}_1- b \vec{\omega}_2}$. In this respect it is convenient to use the invariance of the $\cW_3$ conformal blocks under the exchange $\vec{\omega}_1 \leftrightarrow \vec{\omega}_2$ and to consider the following conformal block: 
\begin{equation}
\label{cbconsidered2b}
\begin{tikzpicture}
[line width=1.2pt]
\draw (-2.5,0)node[left]{$\cB_M (z)$};
\draw (-2.5,0) node[right]{$\mydef$};
\draw (0,0)--(1,0);
\draw (1,0)--(1,1);
\draw (1,0)--(3,0);
\draw (3,0)--(3,1);
\draw (3,0)--(4,0);
\draw (0,0) node[left]{$\Phi^*_L (\infty)$};
\draw (1,1) node[above]{$\Phi_{-b \vec{\omega}_1+s\vec{\omega}_2} (1)$};
\draw (3,1) node[above]{$\Phi_{1211} (z)$};
\draw (4,0) node[right]{$\Phi_R (0)$};
\draw (2,0) node[below]{$M$};
\end{tikzpicture}
\end{equation}
Note that we have kept the fields $\Phi_{R,L}$ un-exchanged as they are general fields.  The computation of the differential equation satisfied by \ref{cbconsidered2b} is strictly similar to the one for \ref{cbconsidered}, the only difference being the fact that $q_1$ gets an opposite sign with respect to the previous case \ref{hq2111}. The resulting sixth order equation has the same pattern of local exponents as in 
\ref{Riem_not}. Their precise values are given by:

\begin{eqnarray}
\label{alphas2}
\alpha_1 &=&\frac13 \left(2 a_{R_2} b + a_{R_1} b\right),
\quad
\alpha_2 =\frac13 \left(3 - a_{R_2} b + a_{R_1} b + 3 b^2\right),  
\quad   
\alpha_3 =\frac13 \left(6 - a_{R_2} b - 2 a_{R_1} b + 6 b^2\right) ,  
\nonumber \\
\label{betas2}
\beta_1 &=& \frac13 \left(3+2 b^2 - b s\right),
\quad \beta_2 =\frac13 \left(6+8 b^2 - b s\right), 
\quad
\beta_3 = \frac13 \left(- b^2 + 2 b s\right)  ,\nonumber \\
\label{gammas2}
\gamma_1 &=& \frac13 \left(-6+2 a_{L_1}b + a_{L_2}b - 8 b^2\right),\,
\gamma_2 = \frac13 \left(-3 - a_{L_1} b +  a_{L_2} b - 5 b^2\right),\,
\gamma_3 = \frac13 \left(-a_{L_1} b -2 a_{L_2} b - 2 b^2\right), \nonumber \\
&&
\end{eqnarray} 
and correspond to the following fusion rules:
in the $s,t,u$ channels:
\begin{itemize}
\item  $s-$channel:

\begin{eqnarray}
\label{ch12}
\text{Channel 1,2}\;\; (\alpha_1):        \quad 
\vec{\alpha}_M^{(s)} &=& \vec{\alpha}_R- b \,     \vec{\omega}_2 \\
\label{ch22}
\text{Channel 3,4        }\; (\alpha_2):\quad 
      \vec{\alpha}_M^{(s)} &=& \vec{\alpha}_R+ b \, \ll \vec{\omega}_2 - \vec{\omega}_1 \rr   \\
\label{ch342}
\text{Channel 5,6        }\; (\alpha_3):\quad
\vec{\alpha}_M^{(s)} &=& \vec{\alpha}_R+ b \,     \vec{\omega}_1  
\end{eqnarray}

\item $t-$channel:
\begin{eqnarray} 
\label{ch1t2}
\text{Channel 1,2,3}\; (\beta_1):        \quad 
\vec{\alpha}_M^{(t)} &=& \ll - b\; \vec{\omega}_1+s\;\vec{\omega}_2\rr \,-b     \ll \vec{\omega}_1-\vec{\omega}_2\rr \\
\label{ch2t2}
\text{Channel  4}\;\; (\beta_2):\quad 
      \vec{\alpha}_M^{(t)} &=& \ll - b\; \vec{\omega}_1+s\;\vec{\omega}_2\rr+ b \, \vec{\omega}_1     \\
\label{ch34t2}
\text{Channel 5,6        }\; (\beta_3):\quad
\vec{\alpha}_M^{(t)} &=&\ll - b\; \vec{\omega}_1+s\;\vec{\omega}_2\rr- b \,     \vec{\omega}_2  
\end{eqnarray}
\item   $u-$channel:

\begin{eqnarray}
\label{ch1u2}
\text{Channel 1,2}\;\; (\gamma_1):        \quad 
\vec{\alpha}_M^{(u)} &=& \vec{\alpha}_L^*- b \,     \vec{\omega}_2 \\
\label{ch2u2}
\text{Channel  3,4        }\; (\gamma_2):\quad 
      \vec{\alpha}_M^{(u)} &=& \vec{\alpha}_L^*+ b \, \ll \vec{\omega}_2 - \vec{\omega}_1 \rr   \\
\label{ch34u2}
\text{Channel 5,6       }\; (\gamma_3):\quad
\vec{\alpha}_M^{(u)} &=& \vec{\alpha}_L^*+ b \,     \vec{\omega}_1  
\end{eqnarray}

\end{itemize}

The above values are consistent with the expected fusion rules and the analysis of the degeneracy pattern is strictly analogous to the one done above for the case \ref{cbconsidered}.
\section{Summary and discussion}
\label{section.05.summary.comments}

The main motivation of our study lies in the fact that, in $\cW_{N\geq 3}$ theories, the general matrix 
element of a primary field between two descendant states is not expressed solely in terms of the primary 3-point function 
but involves also an infinite set of new independent basic matrix elements. This  greatly limits the available information on correlation functions. Another manifestation of this is that the AGT correspondence \cite{agt09, wyl09} for $\cW_N$ theories allows to construct matrix elements only for the fields with highest-weights proportional either to $\omega_1$ or to $\omega_{N-1}$ fundamental weights of $\slN$. In this case, the correspondence between 2-dimensional conformal field theory and 4-dimensional 
supersymmetric gauge theories, as proposed in \cite{agt09}, is available, and $\cW_N$ 
conformal blocks are equal to Nekrasov instanton partition functions \cite{wyl09}. See \cite{bonelli2012vertices, gl14, pom14I, pom14II, furlan2015some} for recent works
towards a more general analysis.

In this paper we focused our attention on the field $\Phi_{-b \vec{\omega}_1+s \vec{\omega}_2}$ of the $\cW_3$ Toda conformal field theory.  We showed that this field  is a second level semi-degenerate field and we found the corresponding null-vector conditions (\ref{nullvector2}). These conditions allow for the computation of all except one matrix elements involving this field: the basis elements (\ref{me_basis}) with $p=2,3,\cdots$ can indeed be computed as a function of the matrix element (\ref{me_basis}) with $p=1$. The (\ref{p2p1}) is an example  of such relations. Moreover we derived the differential equation obeyed by the conformal block containing a fully-degenerate fundamental fields $\Phi_{-b \vec{\omega}_1}$, the semi-degenerate field $\Phi_{-b \vec{\omega}_1+s \vec{\omega}_2}$ and two general fields $\Phi_{R}$ and $\Phi_{L}$. We computed the local exponents of this Fuchsian equation and we related the corresponding multiplicities to the number of undetermined matrix elements (\ref{me_basis}). Interestingly we also argued that the field $\Phi_{-2b \vec{\omega}_1+s \vec{\omega}_1}$ is a semi-degenerate field at level 3 and the associated matrix elements (\ref{me_basis}) with $p=3,4,\cdots$ can be computed, via the differential equation, in terms of the ones with $p=1,2$.  

Our results demand further investigations. To begin with, it would be interesting to find the monodromy group associated to the sixth-order differential systems. This would allow the definition of local correlation functions, to compare with the ones computed in \cite{fl08} by completely different methods, and the determination of $\cW_3$ structure constants that are, at the present, unknown. 
Moreover, it would be interesting to understand how to recover our results for the  semi-degenerate level-2 fields using the AGT correspondence. In this respect, the case of central charge $c=2$ can be tackled with the methods proposed in \cite{Gavrilenko2016}.

\appendix

%SECTION.APPENDIX.A
\section{Details of the derivation}
\label{detailderiv}

Besides the conformal block $ \cB_M\ll L, 2, 1, R \rr  (\{z_X\})$,  we need to consider the two functions:
\begin{eqnarray}
 W_{-1}^{(2)}\cB_M\ll L, 2, 1, R \rr  (\{z_X\})&=&\left< \Phi^{*}_{L}(z_L) \left[W_{-1}\Phi_{-b\vec{\omega}_1+s\vec{\omega}_2}\right](z_2)  \Phi_{-b\vec{\omega}_{1}}(z_1)  \Phi_{R}(z_R)   \right> \;,\nonumber \\
\left[W_{-1}^{2}\right]^{(2)}\cB_M\ll L, 2, 1, R \rr (\{z_X\})&=& \left< \Phi^{*}_{L}(z_L) \left[W_{-1}^2\Phi_{-b\vec{\omega}_1+s\vec{\omega}_2}\right](z_2)  \Phi_{-b\vec{\omega}_{1}}(z_1)  \Phi_{R}(z_R)\right>\;.\nonumber \\
&&  
\end{eqnarray} 
Analogously to (\ref{global_conformal}), the global conformal invariance fixes the coordonate dependence:

\begin{eqnarray}
W_{-1}^{(2)} \cB_M \ll L, 2, 1, R \rr (\{z_X\}) &=&\left(z_L-z_1\right)^{-2  h_1} \left(z_L-z_R\right)^{h_1-h_R+h_2+1-h_L}  \times\nonumber \\
&\times& \left(z_L-z_2\right)^{h_1+h_R-h_2-1-h_L}\left(z_2-z_R\right)^{-h_1-h_R-h_2-1+h_L}
 H(z),\nonumber  \\
 \left[W_{-1}^2\right]^{(2)} \cB_M \ll L, 2, 1, R \rr (\{z_X\}) &=&\left(z_L-z_1\right)^{-2  h_1} \left(z_L-z_R\right)^{h_1-h_R+h_2+2-h_L} \times\nonumber \\
&\times&  \left(z_L-z_2\right)^{h_1+h_R-h_2-2-h_L} \left(z_2-z_R\right)^{-h_1-h_R-h_2-2+h_L}
 H_1(z),\nonumber
 \label{global_conformal_2}
\end{eqnarray}
where
\be
H(z)\mydef W_{-1}^{(2)}\mathcal{B}_M(z)\;, \qquad H_1(z)\mydef \left[ W_{-1}^2\right]^{(2)} \mathcal{B}_M(z)\;,
\ee
and $z$ is given in (\ref{anharmonic_factor}).  We recall that the values of $(h_1, q_1)$ and $(h_2,q_2)$, characterizing the fields at position $z_{1}$ and $z_2$, are given respectively in (\ref{hq2111}) and in (\ref{quantumnumberssemi}). 
Using the \eqref{nullvector2} for the semi-degenerate field at $z_2$, one has the relation
\be \label{blockG}
 c^{\text{sing}}_{\{2;\emptyset\}}  L_{-2}^{(2)}\cB_M(z) +
c^{\text{sing}}_{\{\emptyset;2\}}  W_{-2}^{(2)}\cB_M(z)   +
c^{\text{sing}}_{\{1,1;\emptyset\}}  \left[L_{-1}^2\right]^{(2)}\cB_M(z) + 
c^{\text{sing}}_{\{1;1\}}  L_{-1}^{(2)}H(z) + H_{1}(z) =0.
%\sum_{\substack{
%I
%\\ 
%|I| = 2}} c^{\text{sing}}_{I}  \mathcal{L}_{I}^{(2)} \cB_M(z)  =0\;,
\ee
where $ c_{I}^{\text{sing}}$ are the components of the vector $c^{\text{sing}}$ given in (\ref{csing}). 
\noindent The functions $L_{-2}^{(2)}\cB_M(z)$ and $\left[L_{-1}^2\right]^{(2)}\cB_M(z)$, related to conformal blocks involving pure  Virasoro descendants, can be expressed in terms of  differential operators acting  on
$\cB_{M}(z)$:  
\begin{align}
&L_{-2}^{(2)}\cB_M(z)=\;\frac{(z-2) z}{z-1} \cB_M'(z)+\bigg[h_1+2 h_R+h_2-h_L+\frac{h_1}{(z-1)^2}\bigg]\cB_M(z) \;,\\
&\left[L_{-1}^2\right]^{(2)}\cB_M(z)=\; z^2 \cB_M''(z)+ 2z \left(h_1+h_R+h_2-h_L+1\right)  \cB_M'(z)+\nonumber\\
&\hspace{4cm} (h_1+h_R +h_2-h_L)\left(h_1+h_R+h_2-h_L+1\right)  \cB_M(z)\;,
 \end{align}
 while the function $L_{-1}^{(2)}H(z)$ is easily expressed as
 \begin{equation}
 L_{-1}^{(2)}H(z)=\; -\left(h_1+h_R+h_2-h_L+1\right) H(z)-z H'(z)\;.
 \end{equation}
Less direct is the computation of $ W_{-2}^{(2)}\cB_M(z)$ term.  
Using five $\cW$-Ward identities for $\cB_M(z)$ together with the three null-vector conditions (\ref{Level.1.identity}), (\ref{Level.2.identity}) and (\ref{Level.3.identity}) for the field $\Phi_{-b \vec{\omega}_1}(z_1)$   allows to express $ W_{-2}^{(2)}\cB_M(z)$  in terms of $\cB_M(z)$ and $H(z)$. The resulting expressions is

\begin{align}
\label{W2B}
W_{-2}^{(2)}\cB_M(z)=\;& -2 H(z) - \frac{12 q_1 z^2 \cB_M''(z)}{h_1 \left(5 h_1+1\right)}-\frac{3 q_1 z \left(h_1 (9 z-7)+5 z-3\right) \cB_M'(z)}{h_1 \left(5 h_1+1\right) (z-1)}-
\nonumber\\&
\bigg[\frac{q_1 \left(5 h_1-11 h_1 z-6 h_L z+7 z-1\right)}{\left(5 h_1+1\right) (z-1)}+
\frac{6 q_1 \left(h_1 h_R+h_R-h_2 z-h_1 h_2 z-h_L z\right)}{h_1 \left(5 h_1+1\right) (z-1)}
\nonumber\\&
+  \frac{2 h_1 q_1 \left(11 z^2-16 z+5\right)}{\left(5 h_1+1\right) (z-1)^2}+
\frac{6 h_2 (h_1+1) q_1 (z-2) z}{h_1 \left(5 h_1+1\right) (z-1)^2}+q_R+q_2-q_L\bigg] \cB_M(z)\;.
\end{align}

Finally we have to express the function  $H(z)$ and $H_1(z)$ in terms of the differential operator acting on $\cB_M(z)$. This can be done by using the Ward identities (\ref{WW: ward2_1})-(\ref{WW: ward2_5}). We obtain the following two identities:
\be\label{blockH}
\frac{1}{(z-1)^2 z^2} H(z)+g_0 \cB_M(z)+g_1 \cB_M'(z)+g_2 \cB_M''(z)+g_3 \cB_M^{'''}(z)=0\;, 
\ee
and 
\be\label{blockF} 
\frac{1}{(z-1)^2 z^2} H_1(z)+ \tilde{g}_0 \cB_M(z)+\tilde{g}_1 \cB_M'(z)+l_0 H(z)+l_1 H'(z)+l_2 H''(z)  +l_3 H^{'''}(z)=0\;.
\ee
The coefficients in the above equations read respectively: 
\begin{align}
g_0=\;&\frac{3 \left(3 h_1-1\right) h_Rq_1(2-3 z)}{2 h_1\left(5 h_1+1\right) (z-1)^2 z^3}-\frac{ \left(q_R(z-1)^2-q_Lz (z-1)^2-q_2(2 z-1) z\right)}{(z-1)^3 z^3}
\nonumber\\
&+\frac{3 h_2q_1 \left(h_1(9 z-7)-3 z-3\right)}{2 h_1\left(5 h_1+1\right) (z-1)^3 z^2}+\frac{3 h_Lq_1 \left(h_1(10 z-7)+2 z-3\right)}{2 h_1\left(5 h_1+1\right) (z-1)^2 z^2}
\nonumber\\
&+\frac{q_1  \left(h_1(31-40 z)-8 z+11\right)}{2 \left(5 h_1+1\right) (z-1)^2 z^2}\;,\\
g_1=\;&-\frac{3 q_1\left(44 z^2-53 z+12\right)}{2 \left(5 h_1+1\right) (z-1)^2 z^2}-\frac{12 q_1\left(h_R(z-1)-h_2z\right) }{h_1\left(5 h_1+1\right) (z-1)^2 z^2}
\nonumber\\
&-\frac{3 q_1\left(20 z^2-25 z+4\right) }{2 h_1\left(5 h_1+1\right) (z-1)^2 z^2}+\frac{12 h_Lq_1}{h_1\left(5 h_1+1\right) (z-1) z}\;,\\
g_2=\;&\frac{12 q_1(1-z) (5 z-3) }{h_1\left(5 h_1+1\right) (z-1)^2 z}\;,\qquad
g_3=\frac{16 q_1(1-z)}{h_1\left(h_1+1\right) \left(5 h_1+1\right) (z-1)}\;,
\end{align}
and 
\begin{align}
%f_0=\;&\frac{1}{(z-1)^2 z^2}\;&
%\\
\tilde{g}_0=\;&\frac{9 q_1q_2 \left(h_1(9 z-8) -3 z-8\right)}{2 h_1\left(5 h_1+1\right) (z-1)^4 z}+\frac{\left(h_1+h_R\right) \kappa (1-2 z) }{(z-1)^3 z^2}-
\frac{h_L\kappa (1-2 z)}{(z-1)^3 z^2}\nonumber\\
&-\frac{h_2\kappa \left(z^2-3 z+1\right) }{(z-1)^4 z^2}\;,
\\
\tilde{g}_1=\;&\frac{36 q_1w_3}{h_1\left(5 h_1+1\right) (z-1)^3}+\frac{\kappa (1-2 z)}{(z-1)^3 z}\;,
\\
l_0=\;& \frac{q_1\left(h_1z (31-40 z)+9 h_R(2-3 z)\right)}{2 \left(5 h_1+1\right) (z-1)^2 z^3}
-\frac{9 \left(h_2+1\right) q_1(z+1)}{2 h_1\left(5 h_1+1\right) (z-1)^3 z^2}
\nonumber\\
&-\frac{ \left(q_R(z-1)^2-q_Lz (z-1)^2+q_2z (1-2 z)\right)}{(z-1)^3 z^3}+\frac{3 h_2q_1(9 z-7)}{2 \left(5 h_1+1\right) (z-1)^3 z^2}
\nonumber\\
&-\frac{q_1\left(4 z^2-23 z+16\right)}{\left(5 h_1+1\right) (z-1)^3 z^2}+\frac{3 q_1\left(h_R(3 z-2)+h_L(2 z-3) z+h_1h_L(10 z-7) z\right)}{2 h_1\left(5 h_1+1\right) (z-1)^2 z^3}\;,
\\
l_1=\;&\frac{12 q_1\left(h_Lz-h_R\right)}{h_1\left(5 h_1+1\right) (z-1) z^2}-
\frac{3 q_1(44 z^2-53z+12)}{2 \left(5 h_1+1\right) (z-1)^2 z^2}\nonumber\\
&-\frac{3 q_1\left(20 z^2-33 z+4\right)}{2 h_1\left(5 h_1+1\right) (z-1)^2 z^2}+\frac{12 h_2w_1}{h_1\left(5 h_1+1\right) (z-1)^2 z}\;,\\
l_2=\;& \frac{12 q_1(3-5 z)}{h_1\left(5 h_1+1\right) (z-1) z}\;,
\qquad
l_3=-\frac{16 q_1}{h_1\left(h_1+1\right) \left(5 h_1+1\right)}\;.
\end{align}

Using \eqref{blockH} and  \eqref{blockF}, we are able to express $H(z)$ and $H_1(z)$ in terms of $\cB_M(z)$ and its derivatives (up to 
6th order). Using these results in \eqref{blockG},  we finally get the 6th order differential equation for 
$\cB_M(z)$. 

%SECTION.APPENDIX.C

%SECTION.ACK
\section*{Acknowledgements}
We thank the Institut Henri Poincare, Paris, where this work was initiated, and the Poncelet Laboratory (Moscow) 
where this work was ended, for excellent hospitality and financial support. 
The work of V.B. was performed at the Landau Institute for Theoretical Physics,
with the financial support from the Russian Science Foundation (Grant No.14-50-00150). 
We greatly thank O. Foda for contributions to the early stages of this project. We thank P. Gavrylenko,  N. Iorgov, Y. Ikhlef, Y. Matsuo and S. Ribault for discussions.

\bibliographystyle{JHEP}
\bibliography{bibnonwy.bib}

\providecommand{\href}[2]{#2}\begingroup\raggedright\begin{thebibliography}{10}

\bibitem{zam85}
A.~B. Zamolodchikov, \emph{{Infinite Additional Symmetries in Two-Dimensional
  Conformal Quantum Field Theory}},
  \href{http://dx.doi.org/10.1007/BF01036128}{\emph{Theor. Math. Phys.} {\bf
  65} (1985) 1205--1213}.

\bibitem{fz86b}
V.~A. Fateev and A.~B. Zamolodchikov, \emph{{Conformal quantum field theory
  models in two dimensions having $Z_3$ symmetry}}, {\emph{Nucl. Phys.} {\bf
  B280} (1987) 644--660}.

\bibitem{bs95}
P.~Bouwknegt and K.~Schoutens, \emph{{W symmetry}}, {\emph{Adv.Ser.Math.Phys.}
  {\bf 22} (1995) 1--875}.

\bibitem{bpz84}
A.~A. Belavin, A.~M. Polyakov and A.~B. Zamolodchikov, \emph{{Infinite
  conformal symmetry in two-dimensional quantum field theory}}, {\emph{Nucl.
  Phys.} {\bf B241} (1984) 333--380}.

\bibitem{fl07c}
V.~A. Fateev and A.~V. Litvinov, \emph{{Correlation functions in conformal Toda
  field theory I}}, {\emph{JHEP} {\bf 11} (2007) 002},
  [\href{http://arxiv.org/abs/0709.3806}{{\tt 0709.3806}}].

\bibitem{fl08}
V.~A. Fateev and A.~V. Litvinov, \emph{{Correlation functions in conformal Toda
  field theory II}}, {\emph{JHEP} {\bf 01} (2009) 033},
  [\href{http://arxiv.org/abs/0810.3020}{{\tt 0810.3020}}].

\bibitem{kms10}
S.~Kanno, Y.~Matsuo and S.~Shiba, \emph{Analysis of correlation functions in
  toda theory and the alday-gaiotto-tachikawa-wyllard relation for s u (3)
  quiver}, {\emph{Physical Review D} {\bf 82} (2010) 066009}.

\bibitem{agt09}
L.~F. Alday, D.~Gaiotto and Y.~Tachikawa, \emph{{Liouville Correlation
  Functions from Four-dimensional Gauge Theories}}, {\emph{Lett. Math. Phys.}
  {\bf 91} (2010) 167--197}, [\href{http://arxiv.org/abs/0906.3219}{{\tt
  0906.3219}}].

\bibitem{wyl09}
N.~Wyllard, \emph{{A(N-1) conformal Toda field theory correlation functions
  from conformal N = 2 SU(N) quiver gauge theories}}, {\emph{JHEP} {\bf 11}
  (2009) 002}, [\href{http://arxiv.org/abs/0907.2189}{{\tt 0907.2189}}].

\bibitem{mironov2010agt}
A.~Mironov and A.~Morozov, \emph{On agt relation in the case of u (3)},
  {\emph{Nuclear physics B} {\bf 825} (2010) 1--37}.

\bibitem{Belavin:2011pp}
V.~Belavin and B.~Feigin, \emph{{Super Liouville conformal blocks from N=2
  SU(2) quiver gauge theories}},
  \href{http://dx.doi.org/10.1007/JHEP07(2011)079}{\emph{JHEP} {\bf 07} (2011)
  079}, [\href{http://arxiv.org/abs/1105.5800}{{\tt 1105.5800}}].

\bibitem{aflt10}
V.~A. Alba, V.~A. Fateev, A.~V. Litvinov and G.~M. Tarnopolsky, \emph{{On
  combinatorial expansion of the conformal blocks arising from AGT
  conjecture}}, \href{http://dx.doi.org/10.1007/s11005-011-0503-z}{\emph{Lett.
  Math. Phys.} {\bf 98} (2011) 33--64},
  [\href{http://arxiv.org/abs/1012.1312}{{\tt 1012.1312}}].

\bibitem{epss11}
B.~Estienne, V.~Pasquier, R.~Santachiara and D.~Serban, \emph{{Conformal blocks
  in Virasoro and W theories: Duality and the Calogero-Sutherland model}},
  {\emph{Nucl.Phys.} {\bf B860} (2012) 377--420},
  [\href{http://arxiv.org/abs/1110.1101}{{\tt 1110.1101}}].

\bibitem{bb11}
A.~Belavin and V.~Belavin, \emph{{AGT conjecture and Integrable structure of
  Conformal field theory for c=1}},
  \href{http://dx.doi.org/10.1016/j.nuclphysb.2011.04.014}{\emph{Nucl.Phys.}
  {\bf B850} (2011) 199--213}, [\href{http://arxiv.org/abs/1102.0343}{{\tt
  1102.0343}}].

\bibitem{BEFS16}
V.~{Belavin}, B.~{Estienne}, O.~{Foda} and R.~{Santachiara}, \emph{{Correlation
  functions with fusion-channel multiplicity in $\{$mathcal$\{$W$\}$$\}$\_3
  Toda field theory}},
  \href{http://dx.doi.org/10.1007/JHEP06(2016)137}{\emph{Journal of High Energy
  Physics} {\bf 6} (June, 2016) 137},
  [\href{http://arxiv.org/abs/1602.03870}{{\tt 1602.03870}}].

\bibitem{bs92}
P.~Bouwknegt and K.~Schoutens, \emph{{W symmetry in conformal field theory}},
  1993.

\bibitem{bw92}
P.~Bowcock and G.~M.~T. Watts, \emph{{Null vectors of the W(3) algebra}},
  {\emph{Phys. Lett.} {\bf B297} (1992) 282--288},
  [\href{http://arxiv.org/abs/hep-th/9209105}{{\tt hep-th/9209105}}].

\bibitem{wattsdet}
G.~Watts, \emph{Determinant formulae for extended algebras in two-dimensional
  conformal field theory},
  \href{http://dx.doi.org/http://dx.doi.org/10.1016/0550-3213(89)90548-8}{\emph{Nuclear
  Physics B} {\bf 326} (1989) 648 -- 672}.

\bibitem{furlan2015some}
P.~Furlan and V.~Petkova, \emph{On some 3-point functions in the $ w\_4 $ cft
  and related braiding matrix}, {\emph{arXiv preprint arXiv:1504.07556} (2015)
  }.

\bibitem{yoshida1987fuchsian}
M.~Yoshida, \emph{Fuchsian differential equations}.
\newblock Springer, 1987.

\bibitem{bpt92}
Z.~Bajnok, L.~Palla and G.~Takacs, \emph{{A(2) Toda theory in reduced WZNW
  framework and the representations of the W algebra}}, {\emph{Nucl. Phys.}
  {\bf B385} (1992) 329--360}, [\href{http://arxiv.org/abs/hep-th/9206075}{{\tt
  hep-th/9206075}}].

\bibitem{bw93}
P.~Bowcock and G.~M.~T. Watts, \emph{{Null vectors, three point and four point
  functions in conformal field theory}}, {\emph{Theor. Math. Phys.} {\bf 98}
  (1994) 350--356}, [\href{http://arxiv.org/abs/hep-th/9309146}{{\tt
  hep-th/9309146}}].

\bibitem{wat94}
G.~M.~T. Watts, \emph{{Fusion in the W(3) algebra}}, {\emph{Commun. Math.
  Phys.} {\bf 171} (1995) 87--98},
  [\href{http://arxiv.org/abs/hep-th/9403163}{{\tt hep-th/9403163}}].

\bibitem{bonelli2012vertices}
G.~Bonelli, A.~Tanzini and J.~Zhao, \emph{Vertices, vortices \& interacting
  surface operators}, {\emph{Journal of High Energy Physics} {\bf 2012} (2012)
  1--22}.

\bibitem{gl14}
J.~Gomis and B.~Le~Floch, \emph{{M2-brane surface operators and gauge theory
  dualities in Toda}},  \href{http://arxiv.org/abs/[1407.1852]}{{\tt
  [1407.1852]}}.

\bibitem{pom14I}
V.~Mitev and E.~Pomoni, \emph{{Toda 3-Point Functions From Topological
  Strings}}, {\emph{JHEP} {\bf 06} (2015) 049},
  [\href{http://arxiv.org/abs/1409.6313}{{\tt 1409.6313}}].

\bibitem{pom14II}
M.~Isachenkov, V.~Mitev and E.~Pomoni, \emph{{Toda 3-Point Functions From
  Topological Strings II}},  \href{http://arxiv.org/abs/[1412.3395]}{{\tt
  [1412.3395]}}.

\bibitem{Gavrilenko2016}
P.~{Gavrylenko} and A.~{Marshakov}, \emph{{Exact conformal blocks for the
  W-algebras, twist fields and isomonodromic deformations}},
  \href{http://dx.doi.org/10.1007/JHEP02(2016)181}{\emph{Journal of High Energy
  Physics} {\bf 2} (Feb., 2016) 181},
  [\href{http://arxiv.org/abs/1507.08794}{{\tt 1507.08794}}].

\end{thebibliography}\endgroup

\end{document}